\documentclass[preprint,showpacs,preprintnumbers,amsmath,amssymb]{revtex4}
\usepackage{bm}

\begin{document}
\title{Coupled spin-charge drift-diffusion equations for the Rashba model}

\author{V.V. Bryksin}
\affiliation{A.F. Ioffe Physical Technical Institute,
Politekhnicheskaya 26, 194021 St. Petersburg, Russia}

\author{P. Kleinert}
\affiliation{Paul-Drude-Intitut f\"ur Festk\"orperelektronik,
Hausvogteiplatz 5-7, 10117 Berlin, Germany}

\date{\today}

\begin{abstract}
Coupled spin-charge drift-diffusion equations are derived for a
biased two-dimensional electron gas with weak Rashba spin-orbit
interaction. The basic equations formally agree with recent
results obtained for spin-orbit coupled small polarons. It is
shown that effects of an in-plane electric field on a homogeneous
spin system can completely be described by an associated in-plane
magnetic field. Exploiting this analogy, we predict among other
things the electric-field equivalent of the Hanle effect.
\end{abstract}

\pacs{72.25.-b, 72.10.-d, 72.15.Gd}

\maketitle

\section{Introduction}
The prospects of a new generation of electronic devices stimulates
a renewed theoretical and experimental interest in the study of
spin effects in semiconductors. A prerequisite for the design of
semiconductor structures, whose function is based on electron
spin, is a due understanding of spin dynamics and spin-polarized
transport. In this field, the spin-orbit interaction (SOI)
received particular interest since it allows purely electric
manipulation of the electron spin. Many studies refer to a
two-dimensional electron gas (2DEG), in which the Rashba SOI
arises because of the quantum well asymmetry in the perpendicular
direction. The SOI leads to a coupling between spin and charge
degrees of freedom, which offers the possibility of controlling
the spin polarization by an electric field.

The theoretical description of spin phenomena in semiconductors
under the influence of SOI is based on appropriate transport
equations. There are numerous approaches
\cite{PRL_4220,Yu_1,Yu_2,PRL_066603,PRB_085109,PRB_014421} that
rely on a separation of drift-diffusion processes for spin and
charge densities. Strictly speaking, such an approach is
inappropriate, when the SOI has to be accounted for. Spin-charge
coupled drift-diffusion transport equations have been derived from
firm microscopic models for extended states of a 2DEG
\cite{PRB_045317,PRL_226602,PRB_155308,PRB_245210,PRB_193316} and
for the hopping transport of small polarons.~\cite{PRB_235302}
Especially, the approach by Mishchenko et al.
\cite{PRB_045317,PRL_226602} for a 2DEG with Rashba SOI initiated
many interesting
studies.~\cite{PRL_076602,Bazaliy,PRL_236601,PRB_235322}
Unfortunately, variants of these calculations suffer from an
inconsistency because SOI contributions from the collision
integral have been disregarded. The corrections lead to a
cancellation of contributions that has been used in a previous
work \cite{PRL_076602} to erroneously predict propagating coupled
spin-charge waves.

The aim of the present paper is to derive spin-charge coupled
drift-diffusion equations for a 2DEG with Rashba SOI that correct
deficiencies of previous approaches and that describe the
influence of an external in-plane electric field on spin
polarization. With the help of universal macroscopic
drift-diffusion equations, the spatial and temporal evolution of
coupled spin-charge disturbances as well as associated charge
accumulation and magnetization are studied in semiconductor
heterostructures with Rashba SOI. Our basic equations, which are
derived for weak SOI, completely agree with results obtained for
the hopping transport of small polarons.~\cite{PRB_235302} The
field-induced homogeneous spin accumulation as well as the
charge-Hall current are treated. Furthermore, it is shown that for
a homogeneous system, the effect of the electric field on spin
polarization can be completely captured by a fictitious magnetic
field. This analogy between the real applied electric field and an
auxiliary magnetic field is used to predict a number of
interesting electric-field effects on spin. We mention the decay
of a spin polarization by a transverse electric field. Using this
electric-field driven Hanle effect by exchanging the in-plane
magnetic field by an electric field in the measurement set up for
the ordinary Hanle effect, one should be able to alternatively
determine electron and spin lifetimes under steady-state
conditions by varying the electric field strength. Another
application refers to the pseudo charge-Hall
effect~\cite{PRL_156602}, which is induced by circular polarized
light via the creation of a permanent spin magnetization.

\section{Kinetic equations}
The effect of an in-plane electric field ${\bm{E}}$ on coupled
spin-charge excitations of semiconducting electrons in an
asymmetric quantum well can be described by a single-particle
Hamiltonian
\begin{eqnarray}
H_{0}&=&\sum_{\bm{k},\lambda }a_{\bm{k}\lambda }^{\dag}\left[ \varepsilon_{%
\bm{k}}-\varepsilon _{F}\right] a_{%
\bm{k}\lambda }-\sum_{\bm{k},\lambda ,\lambda ^{\prime }}\left(
\hbar{\bm{\omega}}_{
\bm{k}} \cdot {\bm{\sigma }}_{\lambda \lambda ^{\prime }}\right) a_{\bm{k}%
\lambda }^{\dag}a_{\bm{k}\lambda ^{\prime }}\nonumber\\
&-&e{\bm{E}} \sum_{\bm{k},\lambda}\left. \nabla_{\bm \kappa}
a^{\dag}_{\bm{k}-\frac{\bm{\kappa}}{2}\lambda}a_{\bm{k}+
\frac{\bm{\kappa}}{2}\lambda}\right|_{\bm{\kappa}=\bm{0}}
+u\sum\limits_{\bm{k},
\bm{k}^{\prime}}\sum\limits_{\lambda}a_{\bm{k}\lambda}^{\dag}a_{\bm{k}^{\prime}\lambda},
\label{Hamil}
\end{eqnarray}
which includes both the Rashba SOI and the short-range
spin-independent elastic scattering on impurities. The Hamiltonian
is expressed by creation ($a_{\bm{k}\lambda }^{\dag}$) and
annihilation ($a_{\bm{k}\lambda }$) operators that depend on the
vector ${\bm{k}}=(k_x,k_y,0)$ and the spin index $\lambda$. We
introduced the Fermi energy $\varepsilon_F$, the vector of Pauli
matrices ${\bm{\sigma}}$, and the strength $u$ of the short-range
elastic impurity scattering, which is characterized by the
momentum relaxation time $\tau$. The energy $\varepsilon_{\bm{k}}$
of free electrons in the 2DEG and the coupling term of the Rashba
SOI are given by
\begin{equation}
\varepsilon_{\bm{k}}=\frac{\hbar^2\bm{k}^2}{2m},\quad
{\bm{\omega}}_{ \bm{k}}=\frac{\hbar}{m}(\bm{K}\times\bm{k}),\quad
\bm{K}=\frac{m\alpha}{\hbar^2}{\bm{e}}_z, \label{eq2}
\end{equation}
where $m$ denotes the effective mass and $\alpha$ the strength of
the SOI. The central quantity of our approach is the spin-density
matrix $\widehat{f}$, the components of which
\begin{equation}
f_{\lambda^{\prime}}^{\lambda}(\bm{k},\bm{k}^{\prime}\mid
t)=\langle
a^{\dag}_{\bm{k}\lambda}a_{\bm{k}^{\prime}\lambda^{\prime}}\rangle_t,
\label{eq3}
\end{equation}
satisfy kinetic equations, which are derived for the corresponding
physical elements $f={\rm Tr}\widehat{f}$ and ${\bm{f}}={\rm
Tr}{\bm{\sigma}}\widehat{f}$. The four components of the
spin-density matrix depend on two wave vectors
$({\bm{k}}+{\bm{k}}^{\prime})/2\rightarrow{\bm{k}}$ and
$({\bm{k}}-{\bm{k}}^{\prime})\rightarrow{\bm{\kappa}}$, which
allow the description of inhomogeneous charge and spin
distributions. Relying on the Born approximation for the treatment
of elastic impurity scattering and restricting to lowest-order
corrections of the SOI to the collision integral, we obtain the
following Laplace-transformed kinetic equations \cite{PRB_165313}
\begin{equation}
sf-\frac{i\hbar}{m}(\bm{\kappa}\cdot\bm{k})f-\frac{i\hbar}{m}\bm{K}({\bm{f}}\times\bm{\kappa})
+\frac{e{\bm{ E}}}{\hbar}\nabla_{\bm{k}}f
=\frac{1}{\tau}(\overline{f}-f)+f_0, \label{kin1}
\end{equation}
\begin{eqnarray}
&&s{\bm{f}}+2({\bm{\omega}}_{\bm{k}}\times{\bm{f}})
-\frac{i\hbar}{m}(\bm{\kappa}\cdot\bm{k}){\bm{f}}
+\frac{i\hbar}{m}(\bm{K}\times\bm{\kappa})f+\frac{e{\bm{
E}}}{\hbar}\nabla_{\bm{k}}{\bm{f}}\nonumber\\
&&=\frac{1}{\tau}(\overline{{\bm{f}}}-{\bm{f}})+\frac{1}{\tau}
\frac{\partial}{\partial\varepsilon_{\bm{k}}}
\overline{f\hbar{\bm{\omega}}_{\bm{k}}}-\frac{\hbar{\bm{\omega}}_{\bm{k}}}{\tau}
\frac{\partial}{\partial\varepsilon_{\bm{k}}}
\overline{f}+{\bm{f}}_0, \label{kin2}
\end{eqnarray}
with the initial charge and spin distribution $f_0$ and
${\bm{f}}_0$, respectively. An integration over the polar angle
$\varphi$ of the vector ${\bm{k}}$ is indicated by a cross line
over respective quantities and $s$ denotes the Laplace variable
that refers to the time $t$. The second and third term on the
right-hand side of Eq.~(\ref{kin2}) stems from spin contributions
of the collision integral, which have to be taken into account to
guarantee that the spin system correctly approaches the state of
thermodynamic equilibrium. The solution of the coupled
integro-differential Eqs.~(\ref{kin1}) and (\ref{kin2}) is
searched for in the long-wavelength and low-frequency regime.

To illustrate our approach, let us first treat the evolution of
charge-density disturbances at zero SOI. As the inelastic
scattering time $\tau_{\varepsilon}$ is usually much larger than
$\tau$, the quasi-momenta thermalize already at the time scale
$\tau_{\varepsilon}>t>\tau$ ($s\tau\ll 1$). During this period,
the density matrix $(f,{\bm{f}})$ approaches its mean value
$(\overline{f},\overline{{\bm{f}}})$ with respect to the angle
$\varphi$. In the following stage of the evolution
$t>\tau_{\varepsilon}$, which lasts till a characteristic
diffusion time $\tau_d$, the carrier density locally approaches
the equilibrium distribution. Consequently, the behavior of
particles in this time interval can be described by the Fermi
function $n(\varepsilon_{\bm{k}})$ with a Fermi energy that
depends on spatial coordinates ${\bm{r}}$ and time $t$:
$f(\varepsilon_{\bm{k}},{\bm{r}}\mid
t)=n(\varepsilon_{\bm{k}}-\varepsilon_F({\bm{r}},t))$. In this
evolution period, the energy is already thermalized although both
the charge and spin densities remain still inhomogeneous. For weak
perturbations
$\varepsilon_F({\bm{r}},t)=\varepsilon_F+\Delta\varepsilon_F({\bm{r}},t)$,
we obtain for the density fluctuation $\delta
f(\varepsilon_{\bm{k}},{\bm{r}}\mid
t)=-\Delta\varepsilon_F({\bm{r}},t)
dn(\varepsilon_{\bm{k}}-\varepsilon_F)/d\varepsilon_{\bm{k}}$.
This result brings us to the separation ansatz
\begin{equation}
\overline{f}(\varepsilon,{\bm{\kappa}}\mid s)=-F({\bm{\kappa}}\mid
s)\frac{n^{\prime}(\varepsilon)}{dn/d\varepsilon_F}\label{a1}
\end{equation}
for the new unknown function $F({\bm{\kappa}}\mid s)$, where
$n=\int d\varepsilon \rho(\varepsilon)n(\varepsilon)$ with
$\rho(\varepsilon)$ being the density of states of the 2DEG. For
brevity, we write $\varepsilon$ instead of $\varepsilon_{\bm{k}}$
and use a prime to indicate a derivative with respect to
$\varepsilon$. It is in line with this discussion and
Eq.~(\ref{a1}) to replace the drift term accordingly
\begin{equation}
\frac{e}{\hbar}E_x\frac{\partial}{\partial k_x}f \rightarrow
eE_x\frac{\hbar
k}{m}\frac{n^{\prime\prime}}{n^{\prime}}\cos(\varphi)
\overline{f}.\label{a2}
\end{equation}
Adopting these approximations, which express the basic
understanding of the drift-diffusion approach, Eq.~(\ref{kin1}) is
easily solved under the condition of vanishing SOI
(${\bm{K}}={\bm{0}}$). A spectral drift-diffusion equation is
obtained by expanding the solution of Eq.~(\ref{kin1}) with
respect to ${\bm{\kappa}}$ and by integrating over the angle
$\varphi$
\begin{equation}
\left[s+D(k)\kappa^2+i\mu\varepsilon\frac{n^{\prime\prime}}{n^{\prime}}
{\bm{E}}\cdot{\bm{\kappa}}
\right]\overline{f}(\varepsilon,{\bm{\kappa}}\mid s)=f_0,
\end{equation}
where the diffusion coefficient and the mobility are given by
$D(k)=(\hbar k)^2\tau/(2m^2)$ and $\mu=e\tau/m$, respectively. The
final integration over the energy $\varepsilon$ leads to the
well-known drift-diffusion equation
\begin{equation}
\left[s+D\kappa^2-i\mu {\bm{E}}\cdot{\bm{\kappa}}
\right]\overline{f}({\bm{\kappa}}\mid s)=f_0,
\end{equation}
for the charge density $\overline{f}({\bm{\kappa}}\mid s)=\int
d\varepsilon\rho(\varepsilon)\overline{f}(\varepsilon,{\bm{\kappa}}\mid
s)$. The relationship between the diffusion coefficient $D$ and
the mobility $\mu$ is given by the Einstein relation
$\mu=(eD/n)dn/d\varepsilon_F$, which is applicable both for Fermi
and Boltzmann statistics.

Within the framework of the drift-diffusion approach, a similar
approximation can be used for the spin components of the density
matrix, when the Rashba SOI is weak ($\Omega=\omega_k\tau\ll 1$,
which is accessible by tuning the SOI coupling constant $\alpha$
via the shape of the confinement potential). In this case, the
spin relaxation time $\tau_s$ is large so that we can focus on the
time hierarchy $\tau <\tau_{\varepsilon} <\tau_s <\tau_d$. Under
the condition $t>\tau_{\varepsilon}$ but $t/\tau_s$ arbitrary, a
nonequilibrium spin polarization exists on the background of
thermalized carrier energies. Therefore, Eqs.~(\ref{a1}) and
(\ref{a2}) can be used also for the spin contributions in the
kinetic Eqs.~(\ref{kin1}) and (\ref{kin2}). Exploiting these
approximations, the following set of linear equations is obtained
for the components of the spin-density matrix
\begin{eqnarray}
&&\sigma f+i\Omega (q_xf_y-q_yf_x)=R-\frac{2eE\tau}{\hbar
k}\varepsilon
\frac{n^{\prime\prime}}{n^{\prime}}\cos(\varphi)\overline{f} \nonumber\\
&&\sigma f_x+2\Omega\cos(\varphi)f_z-i\Omega
q_yf=R_x+\frac{2}{\gamma}\varepsilon\frac{n^{\prime\prime}}{n^{\prime}}\sin(\varphi)\overline{f}
-\frac{2eE\tau}{\hbar k}\varepsilon
\frac{n^{\prime\prime}}{n^{\prime}}\cos(\varphi)\overline{f}_x
\nonumber\\
&&\sigma f_y+2\Omega\sin(\varphi)f_z+i\Omega
q_xf=R_y-\frac{2}{\gamma}\varepsilon\frac{n^{\prime\prime}}{n^{\prime}}\cos(\varphi)\overline{f}
-\frac{2eE\tau}{\hbar k}\varepsilon
\frac{n^{\prime\prime}}{n^{\prime}}\cos(\varphi)\overline{f}_y
\nonumber\\
&&\sigma
f_z-2\Omega(\cos(\varphi)f_x+\sin(\varphi)f_y)=R_z-\frac{2eE\tau}{\hbar
k}\varepsilon
\frac{n^{\prime\prime}}{n^{\prime}}\cos(\varphi)\overline{f}_z
,\label{eqs}
\end{eqnarray}
with $q_{x,y}=\kappa_{x,y}/k$, $\gamma=k/K$ and
\begin{equation}
\sigma=\sigma_0-i\gamma\Omega(q_x\cos\varphi+q_y\sin\varphi),\quad
\sigma_0=s\tau+1.
\end{equation}
It is assumed that the in-plane electric field ${\bm{E}}$ is
oriented along the $x$ axis. Restricting to lowest-order
contributions in ${\bm{\kappa}}$ and ${\bm{E}}$, we obtain
\begin{equation}
R_x=\overline{f}_x+\tau f_{x0}-\frac{i\hbar K\tau}{m\sigma_0^2}
\frac{(\varepsilon n^{\prime})^{\prime}}{n^{\prime}}\kappa_y
\overline{f},\quad R=\overline{f}+\tau f_0,\quad
R_z=\overline{f}_z+\tau f_{z0},
\end{equation}
\begin{equation}
R_y=\overline{f}_y+\tau f_{x0}+\frac{i\hbar K\tau}{m\sigma_0^2}
\frac{(\varepsilon n^{\prime})^{\prime}}{n^{\prime}}\kappa_x
\overline{f}-\frac{\hbar K\tau}{m\sigma_0}eE \frac{(\varepsilon
n^{\prime\prime})^{\prime}}{n^{\prime}}\overline{f}.
\end{equation}
It is straightforward but cumbersome to solve the equations for
$f$, ${\bm{f}}$, to expand the solution with respect to $\kappa$,
and to calculate the final integral over the angle $\varphi$. What
we obtain by this procedure are spectral drift-diffusion equations
for coupled spin-charge excitations. After integrating over the
remaining energy $\varepsilon$, we get our final result
\begin{equation}
\left[\frac{\partial}{\partial t}-i(\bm{\kappa}\cdot
\mu\bm{E})+D\kappa^2 \right]{F}-\frac{i\hbar}{m}\bm{\kappa}\cdot
\left[\bm{K}\times {\bm{F}}
\right]+\frac{2i\hbar\mu}{e}\left(\bm{\kappa}\cdot\left[\bm{K}
\times\mu\bm{E} \right] \right)(\bm{K}\cdot
{\bm{F}})=0,\label{fun1}
\end{equation}
\begin{eqnarray}
&&%
\left[\frac{\partial}{\partial t}-i(\bm{\kappa}\cdot
\mu\bm{E})+D\kappa^2+\frac{\widehat{A}}{\tau_{s}}
\right]{\bm{F}}+4D\biggl\{\left[\bm{K}\times\left(i\bm{\kappa}
+\frac{\mu}{2D}\bm{E} \right) \right]\times {\bm{F}}\biggl\}
\nonumber\\%
&&%
+\biggl\{%
\frac{2i\hbar\mu}{e}\left(\bm{\kappa}\cdot\left[\bm{K}\times\mu\bm{E}
\right] \right)\bm{K}%
-\frac{2\hbar\mu}{e\tau_s}\left[\bm{K}\times(i\bm{\kappa}+\frac{\mu}{2D}\bm{E})
\right]%
\biggl\}{F}=0,\label{fun2}
\end{eqnarray}
with $A_{xx}=A_{yy}=1$, $A_{zz}=2$, and the spin-scattering time
calculated from $1/\tau_s=4DK^2$. In Eq.~(\ref{fun2}), the
electric field is accounted for via the quasi-chemical potential
($i{\bm{\kappa}}\rightarrow i{\bm{\kappa}}+\mu{\bm{E}}/(2D))$.
These coupled spin-charge drift-diffusion equations are valid for
weak SOI ($\Omega\ll 1$ so that $\tau/\tau_s\ll 1$). What is
interesting is that Eqs.~(\ref{fun1}) and (\ref{fun2}) formally
agree with results, which were recently derived by a different
approach for the hopping transport of small polarons.
\cite{PRB_235302} Spin effects in the latter system exclusively
occur in the weak SOI regime as the lattice constant is much
smaller than typical values of $K^{-1}$. Summarizing this
observation, we point out that the drift-diffusion equations have
an universal character for the Rashba model with weak SOI.

\section{Results and discussion}
The afore-mentioned analogy with the hopping transport provides us
a recipe to transfer results recently obtained for small polarons
\cite{PRB_235302} in a straightforward manner to spin effects of
extended electronic states. We shall not repeat all calculations
already presented in Ref. [\onlinecite{PRB_235302}] but restrict
ourselves to some additional conclusions.

To begin with, let us treat the field-induced homogeneous
(${\bm{\kappa}}={\bm{0}}$) spin accumulation. From
Eq.~(\ref{fun2}), we obtain for the steady-state field-mediated
magnetic moment
\begin{equation}
\overline{{\bm{f}}}=\hbar\mu ({\bm{K}}\times{\bm{E}})n^{\prime}.
\end{equation}
This result expresses the well-known magnetoelectric effect that
was predicted by Edelstein \cite{Edelstein} many years ago. For
its derivation, it was essential to account for spin contributions
on the right-hand side of Eq.~(\ref{kin2}), the origin of which is
the collision integral. If we would neglect these corrections, a
term of the form $i\hbar K\kappa_x \overline{f}/m$ remained
uncompensated in Eq.~(\ref{fun2}), which was used in a recent
paper \cite{PRL_076602} to predict coupled spin-charge waves.
Introducing an electric field via the quasi-chemical potential
\cite{PRB_165313}, this term would lead to a magnetic moment
$\overline{{\bm{f}}}\sim 1/K$, in complete disagreement with well
established results \cite{Edelstein}.

There are numerous other field-mediated spin effects that follow
from Eqs.~(\ref{fun1}) and (\ref{fun2}). As an example, we treat
spin waves that exist in a stripe of a 2DEG oriented parallel to
the electric field (and the $x$-axis). Taking into account
$\tau/\tau_s\ll 1$ and restricting ourselves to long wavelengths
($D\kappa_y^2\tau_s\ll 1$), we obtain the dispersion relation
\begin{equation}
\omega_{1,2}=\frac{3}{2\tau_s}\left[i\pm\sqrt{\tau_s/\tau_E-1}
\right]+D\kappa_y^2\left[i\pm \frac{2}{3\sqrt{\tau_s/\tau_E-1}}
\right],
\end{equation}
in which a rate $1/\tau_E=(\mu E)^2/(9D)$ appears, which is
associated with the electric field. Field-induced damped
oscillations arise at sufficiently high electric field strengths
($\tau_s/\tau_E>1$). For typical parameters $\mu\approx
10^6$~cm$^2$/Vs, $\tau_s\approx 40$~ps, $\tau=0.5$~ps, and
$n=10^{11}$~/cm$^2$, we obtain the condition $E>3.25$~V/cm, which
gives rise to an appreciable current density in the 2DEG. A much
stronger electric field ($\tau_s/\tau_E\gg 1$) drives spin-charge
coupled oscillations with the constant field-dependent frequency
$\omega=\mid \mu E\mid/(2\sqrt{D\tau_s})$. In contrast to the
well-known space-charge waves of free and trapped electrons, this
massive mode with the frequency $\omega=K\mu E$ is independent of
the propagation vector ${\bm{\kappa}}$.

To get further information on spin-charge coupling due to an
electric field, the drift-diffusion Eqs.~(\ref{fun1}) and
(\ref{fun2}) are expressed in spatial coordinates. For the charge
density, we obtain the continuity equation
\begin{equation}
\frac{\partial F({\bm{r}},t)}{\partial t} +{\rm
div}{\bm{j}}({\bm{r}},t)=0,
\end{equation}
with a particle current
\begin{equation}
{\bm{j}}=(\mu{\bm{E}}-D\nabla_{\bm{r}})F+\frac{\hbar}{m}\left[{\bm{K}}\times{\bm{F}}
\right]-\frac{2\hbar\mu^2}{e}K\left[{\bm{K}}\times{\bm{E}}
\right]F_z,\label{charge_current}
\end{equation}
that includes both charge and spin components. Besides the charge
Hall current
\begin{equation}
{\bm{j}}_H=-2\hbar\mu^2 K\left[{\bm{K}}\times{\bm{E}} \right] F_z,
\label{Hallst}
\end{equation}
which arises from a given out-of-plane spin polarization $F_z$,
there appears another spin-related term that is responsible for
the spin-galvanic effect. As the spin is not conserved, the
equation for the spin components $F_{\alpha}$ of the density
matrix cannot be written in the form of a continuity equation.
Rather, we obtain from Eq.~(\ref{fun2})
\begin{equation}
\frac{\partial F_{\alpha}}{\partial
t}+\frac{A_{\alpha}}{\tau_s}F_{\alpha}+2\mu\left(\left[{\bm{K}}\times{\bm{E}}
\right]\times{\bm{F}}
\right)_{\alpha}-\frac{\hbar\mu}{\tau_s}\frac{n^{\prime}}{n}\left[{\bm{K}}\times{\bm{E}}
\right]_{\alpha}F+\frac{\partial J_{i\alpha}}{\partial
r_i}=0,\label{eqq}
\end{equation}
where the spin current $J_{i\alpha}$ is given by
\begin{equation}
{\bm{J}}_{\alpha}=\left(\mu\bm{E}-D\nabla_{\bm{r}}
\right)F_{\alpha}+\delta{\bm{J}}_{\alpha},
\end{equation}
with the spin components
\begin{eqnarray}
&&
\delta{\bm{J}}_0=\frac{\hbar}{m}({\bm{K}}\times{\bm{F}})-\frac{2\hbar
K\tau}{m}({\bm{K}}\times\mu{\bm{E}})F_z,\quad
\delta{\bm{J}}_z=\frac{1}{K\tau_s}\left[{\bm{F}}-\frac{\hbar\tau}{2Dm}
({\bm{K}}\times\mu{\bm{E}})F \right],\nonumber\\
&&
\delta{\bm{J}}_x=-\frac{1}{K\tau_s}\left(F_z{\bm{e}}_x+\frac{\hbar
K^2\tau}{m}F{\bm{e}}_y \right),\quad
\delta{\bm{J}}_y=\frac{1}{K\tau_s}\left(\frac{\hbar
K^2\tau}{m}F{\bm{e}}_x -F_z{\bm{e}}_y\right).\label{spinc}
\end{eqnarray}
Studies of the spin current received a great deal of recent
interest in the
literature.~\cite{PRL_076604,PRB_165313,PRL_266601,PRB_085315,PRB_075319,PRB_075326,PRB_085314}
This interesting discussion is confronted with a serious problem,
namely that the spin is not conserved. As a consequence, different
definitions of the spin current have been put forward in the
literature. According to Eq.~(\ref{spinc}), we obtain for the
spin-Hall current $J_y^z=-2\hbar K^2\mu^2 EF/e$, which is neither
universal~\cite{PRL_126603} nor in line with the result derived
from a more physically motivated definition of the spin-Hall
current.~\cite{PRL_076604,PRB_165313} In addition, there is a
component of the spin current that is completely independent of
the electric field ($J_y^x=-(\hbar K/m)(\tau/\tau_s)F$). This
astonishing result is replaced in an alternative
approach~\cite{PRB_165313} by a spin current that is related to
the initial variation of the spin accumulation and that
disappears, when the latter one reaches its steady-state value. As
the concept of the spin current is not well founded, it seems to
be expedient to avoid its introduction.

To proceed, we further focus on a homogeneous electron gas, for
which the effect of an electric field on the spin polarization can
completely be simulated by an appropriate in-plane magnetic field
of the form
\begin{equation}
{\bm{H}}_{\mathrm{eff}}=\frac{\hbar\mu}{\mu_B}\left[{\bm{K}}\times{\bm{E}}
\right].
\end{equation}
Replacing the electric field by this equivalent magnetic field,
Eq.~(\ref{eqq}) is written as
\begin{equation}
\frac{\partial{\bm{F}}}{\partial
t}+\frac{\widehat{A}}{\tau_s}{\bm{F}} +\frac{2\mu_B}{\hbar}
\left[{\bm{H}}_{\mathrm{eff}}\times{\bm{F}}
\right]-\frac{1}{\tau_s}
\frac{\chi{\bm{H}}_{\mathrm{eff}}}{\mu_B}={\bm{0}},\label{Hform}
\end{equation}
with $\chi =\mu_B^2n^{\prime}$ being the magnetic susceptibility.
The term $\chi{\bm{H}}_{\mathrm{eff}}$ is responsible for the spin
accumulation, whereas the vector product
${\bm{H}}_{\mathrm{eff}}\times{\bm{F}}$ leads to spin precession
around the effective magnetic field ${\bm{H}}_{\mathrm{eff}}$.
This close relationship between spin polarization due to an
electric field and its description by an associated in-plane
magnetic field can be used for the derivation and interpretation
of electric-field effects on spin. As an example, we mention the
rotation of an initial perpendicular homogeneous spin polarization
$F_z(t=0)=F_{z0}$ into the plane of the 2DEG due to Larmor
precession. For a constant electric field oriented along the $x$
axis, we obtain from Eq.~(\ref{Hform})
\begin{equation}
F_x(t)=-\frac{2\mu_B
H_{\mathrm{eff}}}{\hbar}\frac{\sin(\omega_st)}{\omega_s}
\exp\left(-\frac{3t}{2\tau_s} \right)F_{z0},\quad
\omega_s=\sqrt{\left(\frac{2\mu_B H_{\mathrm{eff}}}{\hbar}
\right)^2-\frac{1}{(2\tau_s)^2}},
\end{equation}
with $H_{\mathrm{eff}}=\hbar\mu KE/\mu_B$. This solution
demonstrates that sufficiently high electric fields lead to an
in-plane spin polarization that oscillates with the frequency
$\omega_s\approx 2\mu_BH_{\mathrm{eff}}/\hbar=2\mu EK$.

To further exploit the analogy between electric and magnetic field
effects, let us treat the optical generation and recombination of
a steady-state spin polarization under the additional influence of
a real in-plane magnetic field ${\bm{H}}$ oriented along the $y$
axis. In this case, Eq.~(\ref{Hform}) takes the form
\begin{equation}
\frac{\partial\delta{\bm{F}}}{\partial
t}+\frac{\widehat{A}}{\tau_s}\delta{\bm{F}} +\frac{2\mu_B}{\hbar}
\left[{\bm{B}}\times\delta{\bm{F}}
\right]={\bm{G}}-\frac{\delta{\bm{F}}}{\tau_0},\quad
\delta{\bm{F}}={\bm{F}}-\frac{\chi{\bm{B}}}{\mu_B},\label{Hform1}
\end{equation}
where ${\bm{B}}={\bm{H}}+{\bm{H}}_{\mathrm{eff}}$. The vector
${\bm{G}}$ describes the optical out-of-plane spin generation and
$\tau_0$ is the relaxation time of photogenerated electrons. The
steady-state solution of Eq.~(\ref{Hform1}) is easily obtained
\begin{equation}
\delta
F_z=\frac{\tau_{\perp}G_z}{1+\omega_c^2\tau_{\parallel}\tau_{\perp}},\quad
\frac{1}{\tau_{\parallel}}=\frac{1}{\tau_s}+\frac{1}{\tau_0},\quad
\frac{1}{\tau_{\perp}}=\frac{2}{\tau_s}+\frac{1}{\tau_0},\label{losung}
\end{equation}
with $\omega_c=2\mu_BB/\hbar$. At zero electric field, the
solution describes the depolarization of spin (and hence the
degree of circular polarization of the luminescence) by a
transverse magnetic field, which is known as the Hanle
effect.~\cite{BuchHanle} Combining the measurement of the
zero-field spin generation $G_z$ with the magnetic-field
dependence in Eq.~(\ref{losung}) (Hanle effect), both the electron
lifetime and the spin-relaxation time $\tau_s$ can be
determined.~\cite{Dyakonov85,APL_733,SST_209,APL_022113,PhysE_399,PRB_245319}
According to Eq.~(\ref{losung}), an effect of the same kind exists
also at zero magnetic field (${\bm{H}}={\bm{0}}$) due to an
in-plane electric field. To describe this effect, it is only
necessary to replace the Larmor frequency $2\mu_BB/\hbar$ by
$\omega_c=2\mu EK$ in Eq.~(\ref{losung}). This Hanle effect driven
by a pure in-plane electric field can likewise be used to measure
lifetimes of charge and spin excitations. To the best of our
knowledge, this promising experimental technique has not been
employed hitherto. With respect to the charge transport, there is
the interesting effect that a circular polarized light
illumination induces a pseudo-Hall effect~\cite{PRL_156602} in the
absence of any external magnetic field. A quantitative description
of this Hall contribution is provided by Eqs.~(\ref{Hallst}) and
(\ref{losung}). Another application of the electric-magnetic field
correspondence refers to a modification of recent experiments, in
which the optically induced spin-galvanic effect was
measured.~\cite{Ganichev_gal} Instead of using an external
magnetic field to achieve an in-plane spin polarization necessary
for the occurrence of the effect, one can likewise apply an
in-plane electric field.

\section{Summary}
Based on the density-matrix approach, spin-charge coupled
drift-diffusion equations were derived for extended electronic
states in a 2DEG with weak SOI. The final basic equations agree
with results that were recently obtained for the hopping transport
of spin-polarized polarons.~\cite{PRB_235302} Due to this
correspondence, results on spin transport obtained for localized
and extended states are mutually applicable to each other. In the
course of the derivation, it was found that a consistent treatment
of spin-charge coupling requires a careful consideration of
spin-orbit contributions to the collision integral, which give
rise to a tricky cancellation in transport equations. Disregarding
these corrections is the source of fatal mistakes that plague
former approaches.

Particular emphasis was put on the effect of an electric field on
the spin polarization of a homogeneous 2DEG. It was shown that the
electric field can be replaced by a fictitious in-plane magnetic
field in the drift-diffusion equations for the spin components.
This interpretation of the equations reveals a number of
interesting similarities between spin effects induced by electric
or magnetic fields. From an experimental point of view, most
attractive seems to be the electric-field equivalent of the Hanle
effect, which provides a new possibility to measure lifetimes of
spin and charge excitations by manipulating exclusively an
in-plane external electric field.

\begin{acknowledgments}
Partial financial support by the Deutsche Forschungsgemeinschaft
and the Russian Foundation of Basic Research is gratefully
acknowledged.
\end{acknowledgments}


\begin{thebibliography}{10}

\bibitem{PRL_4220}
M.~E. Flatte and J.~M. Byers, Phys. Rev. Lett. {\bf 84},  4220
(2000).

\bibitem{Yu_1}
Z.~G. Yu and M.~E. Flatte, Phys. Rev. B {\bf 66},  201202  (2002).

\bibitem{Yu_2}
Z.~G. Yu and M.~E. Flatte, Phys. Rev. B {\bf 66},  235302  (2002).

\bibitem{PRL_066603}
I. Zutic, J. Fabian, and S.~D. Sarma, Phys. Rev. Lett. {\bf 88},
066603
  (2002).

\bibitem{PRB_085109}
D. Amico and G. Vignale, Phys. Rev. B {\bf 65},  085109  (2002).

\bibitem{PRB_014421}
I. Martin, Phys. Rev. B {\bf 67},  014421  (2003).

\bibitem{PRB_045317}
E. Mishchenko and B.~I. Halperin, Phys. Rev. B {\bf 68},  045317
(2003).

\bibitem{PRL_226602}
E.~G. Mishchenko, A.~V. Shytov, and B.~I. Halperin, Phys. Rev.
Lett. {\bf 93},
  226602  (2004).

\bibitem{PRB_155308}
A.~A. Burkov, A.~S. Nunez, and A.~H. MacDonald, Phys. Rev. B {\bf
70},  155308
  (2004).

\bibitem{PRB_245210}
F.~X. Bronold, A. Saxena, and D.~L. Smith, Phys. Rev. B {\bf 70},
245210
  (2004).

\bibitem{PRB_193316}
T.~L. Hughes, Y.~B. Bazaliy, and B.~A. Bernevig, Phys. Rev. B {\bf
74},  193316
   (2006).

\bibitem{PRB_235302}
V.~V. Bryksin, H. B\"ottger, and P. Kleinert, Phys. Rev. B {\bf
74},  235302
  (2006).

\bibitem{PRL_076602}
B.~A. Bernevig, X. Yu, and S.~C. Zhang, Phys. Rev. Lett. {\bf 95},
076602
  (2005).

\bibitem{Bazaliy}
Y.~B. Bazaliy, B.~V. Bazaliy, G. G\"untherodt, and S.~S.~P.
Parkin,
  cond-mat/0602517  (2005).

\bibitem{PRL_236601}
B.~A. Bernevig, J. Orenstein, and S.~C. Zhang, Phys. Rev. Lett.
{\bf 97},
  236601  (2006).

\bibitem{PRB_235322}
M.~H. Liu, K.~W. Chen, S.~H. Chen, and C.~R. Chang, Phys. Rev. B
{\bf 74},
  235322  (2006).

\bibitem{PRL_156602}
V.~M. Edelstein, Phys. Rev. Lett. {\bf 95},  156602  (2005).

\bibitem{PRB_165313}
V.~V. Bryksin and P. Kleinert, Phys. Rev. B {\bf 73},  165313
(2006).

\bibitem{Edelstein}
V.~M. Edelstein, Solid State Commun. {\bf 73},  233  (1990).

\bibitem{PRL_076604}
J. Shi, P. Zhang, D. Xiao, and Q. Niu, Phys. Rev. Lett. {\bf 96},
076604
  (2006).

\bibitem{PRL_266601}
E.~M. Hankiewicz, G. Vignale, and M.~E. Flatte, Phys. Rev. Lett.
{\bf 97},
  266601  (2006).

\bibitem{PRB_085315}
P.~Q. Jin and Y.~Q. Li, Phys. Rev. B {\bf 74},  085315  (2006).

\bibitem{PRB_075319}
J. Li and R. Tao, Phys. Rev. B {\bf 75},  075319  (2007).

\bibitem{PRB_075326}
P. Wang, Y.~Q. Li, and X. Zhao, Phys. Rev. B {\bf 75},  075326
(2007).

\bibitem{PRB_085314}
H.~T. Yang and C. Liu, Phys. Rev. B {\bf 75},  085314  (1007).

\bibitem{PRL_126603}
J. Sinova, D. Culcer, Q. Niu, N.~A. Sinitsyn, T. Jungwirth, and
A.~H.
  MacDonald, Phys. Rev. Lett. {\bf 92},  126603  (2004).

\bibitem{BuchHanle}
F. Meier and E. B.~P.~Zakharchenya, {\em Optical Orientation}
(Elsevier,
  Amsterdam, 1984).

\bibitem{Dyakonov85}
M.~I. Dyakonov, V.~A. Marushchak, V.~I. Perel, and A.~N. Titkov,
Zh. Eksp.
  Teor. Fiz. {\bf 90},  1123  (1985).

\bibitem{APL_733}
R.~J. Epstein, D.~T. Fuchs, W.~V. Schoenfeld, P.~M. Petroff, and
D.~D.
  Awschalom, Appl. Phys. Lett. {\bf 78},  733  (2001).

\bibitem{SST_209}
J. F\"urst, H. Pascher, V.~A. Abalmassov, T.~S. Shamirzaev, and
K.~S.
  Zhuravlev, Semicond. Sci. Technol. {\bf 20},  209  (2005).

\bibitem{APL_022113}
G. Itskos, E. Harbord, S.~K. Clowes, E. Clarke, and L.~F. Cohen,
Appl. Phys.
  Lett. {\bf 88},  022113  (2006).

\bibitem{PhysE_399}
O. Maksimov, X. Zhou, M.~C. Tamargo, and N. Samarth, Physica E
{\bf 32},  399
  (2006).

\bibitem{PRB_245319}
V.~F. Motsnyi, P. van Dorpe, W. van Roy, E. Goovaerts, V.~I.
Safarov, G.
  Borghs, and J.~D. Boeck, Phys. Rev. B {\bf 68},  245319  (2003).

\bibitem{Ganichev_gal}
S.~D. Ganichev, E.~L. Ivchenko, V.~V. Bel'kov, S.~A. Tarasenko, M.
Sollinger,
  D. Weiss, W. Wegscheider, and W. Prettl, Nature {\bf 417},  153  (2002).

\end{thebibliography}


\end{document}